\documentclass[journal]{IEEEtran}
\usepackage{amsmath,amssymb,bm,mathtools,color}
\newcommand{\E}{\ensuremath{\operatorname{E}}}
\DeclareMathOperator{\sign}{sgn}
\newcommand*{\matminus}{%
  \leavevmode
  \hphantom{0}%
  \llap{%
    \settowidth{\dimen0 }{$0$}%
    \resizebox{\dimen0 }{\height}{$-$}%
  }%
}
\definecolor{mygray}{gray}{0.7}
\newtheorem{theorem}{Proposition}
\usepackage{array}
\usepackage{cite}
\begin{document}
\title{A Widely Linear Complex Autoregressive Process of Order One}
\author{Adam~M.~Sykulski,~\IEEEmembership{Member,~IEEE,}
        Sofia~C.~Olhede,~\IEEEmembership{Member,~IEEE,}
        and~Jonathan~M.~Lilly,~\IEEEmembership{Senior~Member,~IEEE}
        \thanks{The work of A. M. Sykulski was supported by a Marie Curie International Outgoing Fellowship within the 7th European Community Framework Programme. S. C. Olhede acknowledges funding from EPSRC Responsive mode (EP/L025744/1), EPSRC Leadership Fellowship (EP/I005250/1), and ERC CoG 2015 - 682172 NETS. The work of J. M. Lilly was supported by award 1031002 from the Physical Oceanography program of the United States National Science Foundation.}
\thanks{A. M. Sykulski and S. C. Olhede are with the Department of Statistical Science, University College London, Gower Street, London WC1E 6BT, UK (emails: a.sykulski@ucl.ac.uk, s.olhede@ucl.ac.uk).}
\thanks{J. M. Lilly is with NorthWest Research Associates, PO Box 3027, Bellevue, WA, USA (email: lilly@nwra.com)}
}

\maketitle
\setlength{\arraycolsep}{3pt}
\begin{abstract}
We propose a simple stochastic process for modeling improper or noncircular complex-valued signals. The process is a natural extension of a complex-valued autoregressive process, extended to include a widely linear autoregressive term. This process can then capture elliptical, as opposed to circular, stochastic oscillations in a bivariate signal. The process is order one and is more parsimonious than alternative stochastic modeling approaches in the literature. We provide conditions for stationarity, and derive the form of the covariance and relation sequence of this model. We describe how parameter estimation can be efficiently performed both in the time and frequency domain. We demonstrate the practical utility of the process in capturing elliptical oscillations that are naturally present in seismic signals.
\end{abstract}
\noindent
\copyright 2016 IEEE. Personal use of this material is permitted. Permission from IEEE must be obtained for all other uses, in any current or future media, including reprinting/republishing this material for advertising or promotional purposes, creating new collective works, for resale or redistribution to servers or lists, or reuse of any copyrighted component of this work in other works.
\IEEEpeerreviewmaketitle

\section{Introduction}\label{S:intro}
\IEEEPARstart{C}{omplex}-valued stochastic processes are useful models for parameterizing bivariate signals. Such models are in widespread use in  applications including oceanography~\cite{gonella1972rotary} and functional Magnetic Resonance Imaging~\cite{rowe2005modeling}. The theory for complex-valued representations has been developed both in the context of stochastic processes~\cite{miller1974complex}, as well as for signal processing~\cite{schreier2010statistical}, with notable recent developments in~\cite{adali2011complex} and~\cite{walden2013rotary}. The complex-valued representation is sometimes preferred to the bivariate representation, due to its compactness and interpretability~\cite{olhede2005local}. For example, complex-valued signals can be naturally decomposed into analytic and anti-analytic signals~\cite{olhede2004analytic}, and provide a practical framework for assessing impropriety or noncircularity in a complex signal~\cite{schreier2006generalized}. On the other hand, the bivariate vector representation can provide better physical understanding of the generating mechanism of the modeled process, and its usage is commonplace in the time series community, see e.g.~\cite[Ch. 10--11]{hamilton1994time}.

The complex autoregressive process~\cite{le1988note,picinbono1997second} is a generalization of a real-valued autoregressive process, in which the autoregressive coefficients and noise increments are both complex-valued. Typically, the real and imaginary components of the noise are assumed to be independent and identically distributed~\cite{picinbono1997second}, thus creating statistically isotropic (i.e. circular) oscillations in the signal. The process is therefore said to be statistically {\em circular} or {\em proper}, defined subsequently, as opposed to one that is {\em noncircular} or {\em improper}.

\begin{figure}[t]
\centering{\includegraphics[width=3.4in]{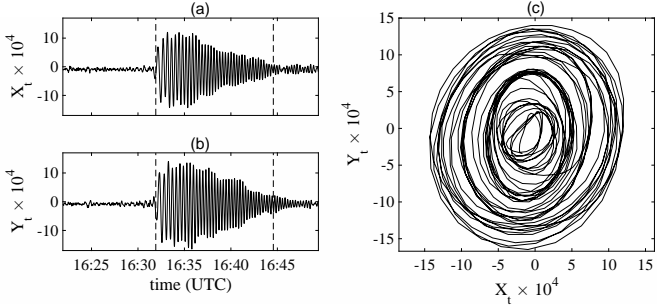}}
\caption{\label{Fig1}Seismic traces from the Feb. 9th, 1991 Solomon Islands earthquake as measured from the Pasadena recording station in California. The radial component, $X_t$, is displayed in (a) and the vertical component, $Y_t$, is displayed in (b). In (c) we display the complex-valued signal, $Z_t=X_t+\mathrm{i}Y_t$, from 16:31:55 to 16:44:35 (UTC), as indicated by the vertical dashed boundaries in (a) and (b).}
\end{figure}

In many real-world observations of complex-valued signals, noncircular or improper structure is expected to be present; examples include seismic traces~\cite{samson1983pure}, oceanographic velocity measurements~\cite{lilly2006wavelet}, and wind measurements~\cite{adali2011complex}. In Fig.~\ref{Fig1}, we display a bivariate signal from a seismic trace of the 1991 Solomon Islands Earthquake, previously studied in~\cite{lilly1995multiwavelet,olhede2003polarization,lilly2011modulated}. In Fig.~\ref{Fig1}(c), the elliptical orbital shape of the oscillations become apparent when viewed in the complex plane, and we will refer to such motion as ``elliptical oscillations." In cases such as these, in which the signal appears improper, a proper process would be a poor choice of model and would fail to summarize important characteristics of the data.

Motivated by this, in this paper we generalize the complex autoregressive processes to a {\em widely linear} complex autoregressive process that is statistically improper or noncircular. The notion of wide linearity was introduced in~\cite{picinbono1995widely}, and we use this concept to relate the complex-valued process $Z_t$ to its previous value $Z_{t-1}$ {\em and} its complex conjugate $Z^\ast_{t-1}$. Our process is order one and hence Markovian, such that $Z_t$ is not dependent on $Z_{t-k}$ (given $Z_{t-1}$) for $k>1$, and is a special case of the widely linear autoregressive moving average (ARMA) model of~\cite{navarro2008arma}. The widely linear ARMA is a general and flexible framework for improper processes that is well understood in the context of signal prediction \cite{navarro2008arma,navarro2013widely}, and signal estimation \cite{navarro2009estimation}, in settings where parameter values are assumed to be known. In practice, these parameters would need to be estimated when modeling real-world signals. 

In this paper, we provide time- and frequency-domain techniques for estimating the parameters of our order one widely linear process, where we will also derive the form of the covariance and relation sequences. A key innovation will be to relate the process to a real-valued bivariate vector autoregressive process, which will provide intuitive understanding of the process generating mechanism, and will simplify the problem to the estimation of five real-valued parameters. We will demonstrate that despite its simplicity, the model we propose can effectively capture the elliptical oscillatory structure present in the seismic signal displayed in Fig.~\ref{Fig1}.

We contrast our widely linear process with the improper complex autoregressive process of~\cite{picinbono1997second,rubin2007simulation}, where the impropriety is created using improper noise. We will demonstrate how the inclusion of a widely linear autoregressive component allows our process to reproduce stochastic elliptical oscillations despite being order one, whereas to generate elliptical oscillations in the framework of~\cite{picinbono1997second}, an order two process is required, as investigated in~\cite{rubin2008kinematics}. We propose our order one process as a simpler and more intuitive model for generating elliptical oscillations in a complex-valued signal.

\section{Background}\label{S:CAR}
Consider a zero-mean complex-valued stationary stochastic process $Z_t$ where $t\in\mathbb{Z}$. The covariance sequence, $s_\tau$, and the relation sequence, $r_\tau$, are defined at lag $\tau\in\mathbb{Z}$ by
\[
s_\tau=\E\{Z_tZ^*_{t+\tau}\},\quad r_\tau=\E\{Z_tZ_{t+\tau}\},
\]
where $\E\{\cdot\}$ denotes expectation and $Z^*_t$ is the complex conjugate of $Z_t$. The process $Z_t$ is said to be proper if
\[
r_\tau=0,\quad \forall\tau\in\mathbb{Z},
\]
and is improper otherwise. A proper process therefore has a relation sequence equal to zero, and is also commonly referred to as a {\em circular} process \cite{schreier2010statistical}. It can be shown that the second-order statistical properties of a proper process are isotropic, such that its distribution is invariant to rotation. Second-order proper processes can be treated much like second-order real-valued processes, in that the second-order properties are fully specified by the covariance sequence.

The improper complex autoregressive process of~\cite{picinbono1997second,rubin2007simulation} is given by
\begin{equation}
Z_t=\sum_{j=1}^pg_jZ_{t-j}+\nu_t,\quad g_j,\nu_t\in\mathbb{C},
\label{eq:doublywhite}
\end{equation}
where $p$ is the order of the process, and where $\nu_t$ is a noise process that is permitted to be noncircular or improper. The coefficients $g_j$ are in general assumed to be complex-valued. An important special case of this process is the proper complex autoregressive process of order one~\cite{le1988note}, which we denote by $Z'_t$ and has three real-valued parameters $\{a,\theta,\sigma^2_\epsilon\}$, and is given by
\begin{equation}
Z'_t = a e^{\mathrm{i}\theta}Z'_{t-1} + \epsilon_t, \quad a\ge0,
\label{eq:CAR}
\end{equation}
where the complex autoregressive coefficient has been expressed in terms of an amplitude $a$ and phase $\theta$, both real-valued. Here $\{\epsilon_t\}$ is a sequence of i.i.d. complex-valued Gaussian noise where the real and imaginary parts are independent, and each has zero mean and variance given by $\sigma^2_\epsilon >0$. The process is stationary if and only if $a<1$, with variance given by $2\sigma^2_\epsilon/(1-a^2)$. In such cases, $a$ is commonly referred to as the damping parameter. For identifiability $a$ is not permitted to be negative, as a negative autoregression is achieved instead when $a>0$ and $\pi/2+2k\pi<\theta<3\pi/2+2k\pi$, where $k\in\mathbb{Z}$. The parameter $\theta\in\mathbb{R}$ is the angle of a rotation of the process at each time step and is usually referred to as the spin parameter~\cite{Arato1962estimation}. The process $Z'_t$ is an example of a proper process, and we will refer back to this process when we construct our widely linear improper process.

The improper complex autoregressive model~\eqref{eq:doublywhite} was generalized to a class of autoregressive moving average (ARMA) models in~\cite{navarro2008arma}, which uses finite-length widely linear filters on both the autoregressive and noise components such that
\begin{equation}
Z_t=\sum_{j=1}^pg_jZ_{t-j}+\sum_{j=1}^ph_jZ^\ast_{t-j}+\sum_{j=0}^qk_j\epsilon_{t-j}+\sum_{j=0}^ql_j\epsilon^\ast_{t-j}.
\label{eq:navarro}
\end{equation}
This more general framework has larger flexibility in modeling improper signals. A challenge however is that as $\{g_j,h_j,k_j,l_j\}$ are complex-valued, then the number of parameters that need to be estimated is large, even for moderate values of $p$ and $q$. In this paper we shall focus on the special case of $p=1$ and $q=0$ in~\eqref{eq:navarro}, thus creating a simple order-one widely linear improper process.

Finally, \cite{picinbono1997second} more generally show that any second-order stationary process can be expressed as a widely linear filter of a complex-proper white noise process $\epsilon_t$ such that 
\begin{equation}
Z_t=\sum_{j=-\infty}^\infty k_j\epsilon_{t-j}+\sum_{j=-\infty}^\infty l_j\epsilon^\ast_{t-j},\quad k_j,l_j,\epsilon_t\in\mathbb{C}.
\label{eq:picinbono}
\end{equation}
This is in essence the complex-analogue to the well-known Wold decomposition for real-valued processes~\cite{wold1954study}, and in general this will be an infinite-order process.

\section{The Widely Linear Complex autoregressive order one process}\label{S:ICAR}
In this section we introduce the widely linear complex autoregressive process of order one. We do this by extending~\eqref{eq:CAR} to a widely linear form for the autoregressive and noise components of the process, such that it is parameterized by seven real-valued parameters. We will subsequently reduce this model to five free parameters, by constraining two parameters, for practical reasons that we shall discuss shortly.

We call our process the widely linear complex autoregressive process of order one, denoted by $Z_t$, which has parameters $\{\lambda,\alpha,\gamma,\phi,\sigma^2_\nu,c_\nu\}$, and is given by
\begin{equation}
Z_t = \lambda e^{\mathrm{i}\alpha}Z_{t-1} + \gamma e^{\mathrm{i}\phi}Z^*_{t-1} + \nu_t, \quad \lambda,\gamma\ge0, \, \label{eq:ICAR}
\end{equation}
where $\{\nu_t\}$ is a sequence of i.i.d. complex-valued Gaussian noise with variance $\sigma^2_\nu = \E\{\nu_t\nu_t^*\}>0$ and relation at lag zero specified by $c_\nu=\E\{\nu_t\nu_t\}\in\mathbb{C}$. The noise process $\nu_t$ is specified by three real-valued parameters as $c_\nu$ is in general complex-valued, therefore our process has seven real-valued parameters. The remaining four parameters $\{\lambda,\alpha,\gamma,\phi\}$ define an iterative relationship between the process $Z_t$ and the previous value $Z_{t-1}$ as well as its conjugate $Z^*_{t-1}$, where $\gamma$ and $\phi$ are respective analogues of the damping parameter, $\lambda$, and the spin parameter, $\alpha$, in this widely linear setting. Further intuition for each of the parameters is gained when we derive covariance sequences in Sections~\ref{S:autocov}.

The widely linear complex autoregressive process of order one is recovered from the ARMA model \eqref{eq:navarro} of~\cite{navarro2008arma} by taking $p=1$ and $q=0$. Furthermore, we combine the widely linear noise term $k_0\epsilon_t+l_0\epsilon^\ast_t$ into the noise process $\nu_t$, which has variance $\sigma^2_\nu=|k_0|^2+|l_0|^2$ and relation at lag zero $c_\nu=2k_0l_0$. For this reason $\nu_t$ is commonly referred to as {\em doubly white noise}~\cite{picinbono1997second}, as it is the pointwise superposition of two complex-proper white noise processes. As we have implicitly chosen the initial phase angle of the noise, we have for parsimony reduced the number of real-valued parameters from the eight used in~\eqref{eq:navarro} (with $p=1$ and $q=0$), to the seven we have defined in~\eqref{eq:ICAR}, without any loss of generality.

\begin{figure}
\centering
\includegraphics[width=2.8in]{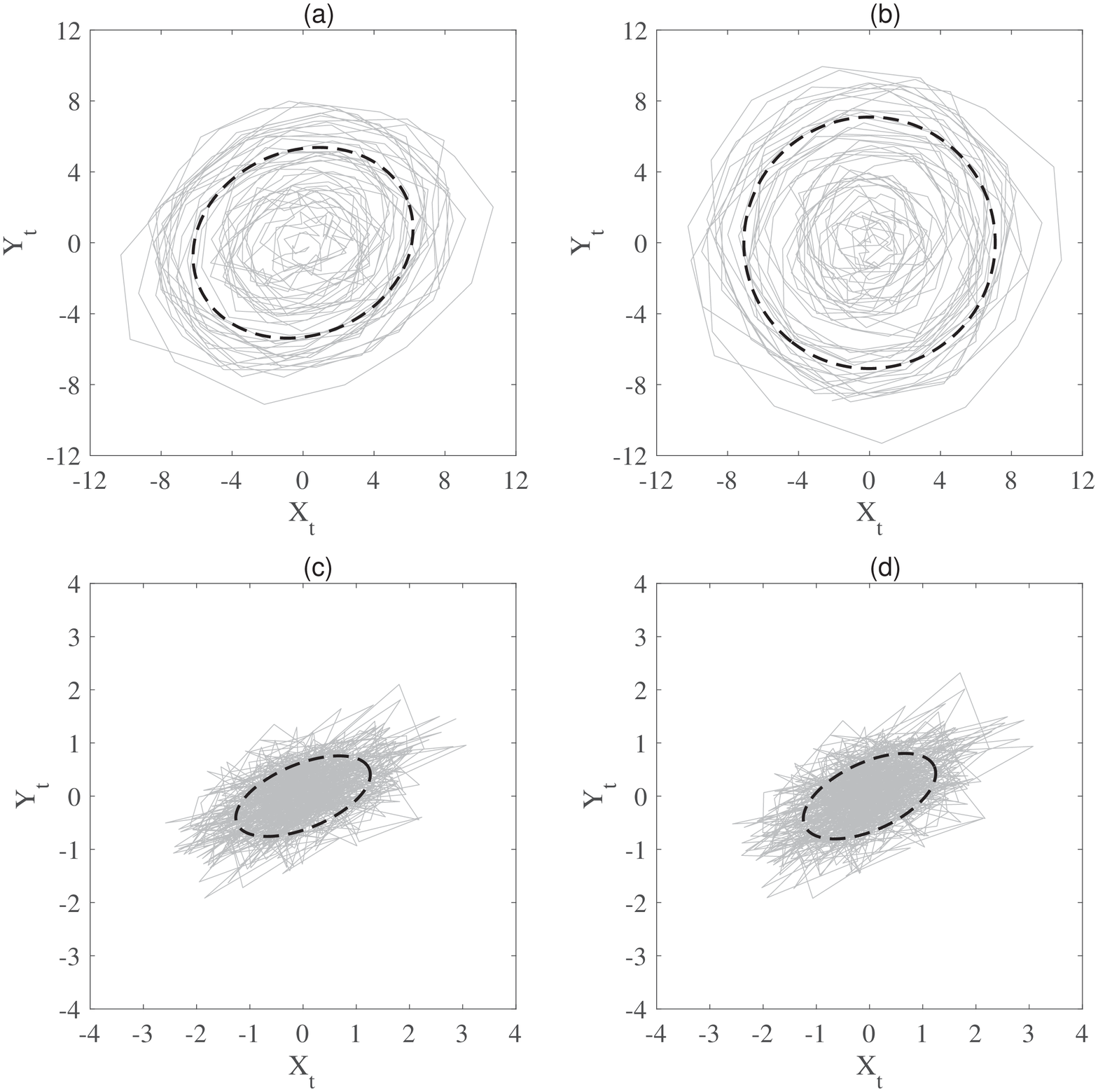}
\caption{\label{FigNew1} Simulated signals of length 512 ({\color{mygray}----}), from the widely linear complex autoregressive process~\eqref{eq:ICAR} in (a) and (c), and from an order one improper process~\eqref{eq:doublywhite} in (b) and (d). In (a) and (b), the parameters are $\lambda=0.99$, $\alpha=\pi/6$, $\sigma^2_\nu=1$, and $c_\nu=\exp(3\pi\mathrm{i}/4)/5$. In (c) and (d), the parameters are $\lambda=0.297$, $\alpha=\pi/6$, $\sigma^2_\nu=1$, and $c_\nu=2\exp(3\pi\mathrm{i}/4)/3$. For the widely linear complex autoregressive process~\eqref{eq:ICAR} in (a) and (c), we additionally set $\gamma=0.099$ and $\phi=-\pi/4$, which are set to zero in (b) and (d) for the model~\eqref{eq:doublywhite}. Expected second moment ellipses for the standard deviation are also given ({- - -}), which are calculated from Section~\ref{S:autocov}.}
\end{figure}

The model~\eqref{eq:ICAR} introduces wide linearity in the autoregressive component of an order one process, and allows the process to map out stochastic elliptical oscillations, as we now demonstrate. In Fig.~\ref{FigNew1}, we contrast realizations from our process with an improper order one process~\eqref{eq:doublywhite} from the framework of~\cite{picinbono1997second} using doubly white noise. The model~\eqref{eq:doublywhite} only has impropriety in the noise component---equivalent to setting $\gamma=0$ in~\eqref{eq:ICAR}. It is clear from the figure that when the autoregressive coefficient $\lambda$ is close to unity (in panels (a) and (b)), then the widely linear complex autoregressive process~\eqref{eq:ICAR} has a tendency to generate elliptical oscillations (panel (a)), whereas an improper order one process~\eqref{eq:doublywhite} generates oscillations that appear to be circular (panel (b)). This is because the motion~\eqref{eq:doublywhite} is largely determined by $Z_t=\lambda e^{\mathrm{i}\alpha}Z_{t-1}$, a deterministic component which specifies a circular oscillation. When $\lambda$ is low (panels (c) and (d)), then the noise term $\nu_t$ dominates both signals. As $\nu_t$ is doubly white noise, then both processes are improper in their distribution, but neither generate elliptical oscillations that resemble the seismic traces seen in Fig~\ref{Fig1}(c)---this can only be generated using~\eqref{eq:ICAR} with a larger autoregressive coefficient, and with $\gamma>0$.

In general, the process~\eqref{eq:ICAR} has seven real-valued parameters, whereas~\eqref{eq:CAR} has three. The additional four parameters in~\eqref{eq:ICAR} are present because each component in~\eqref{eq:CAR}---the deterministic autoregressive component and the noise component---has been given its own improper elliptical structure with two parameters respectively, one to {\em stretch} ($\gamma$ and $|c_\nu|$) and one to {\em rotate} ($\phi$ and $\arg\{c_\nu\}$). Here we have defined $|c_\nu|$ as the complex modulus of $c_\nu$, and $\arg\{c_\nu\}$ as the complex argument or phase. In practical problems the four parameters $\{\gamma,\phi,|c_\nu|,\arg\{c_\nu\}\}$ will be difficult to identify simultaneously from observed signals. For this reason we reduce the widely linear process~\eqref{eq:ICAR} to five free parameters, effectively making two parameters redundant, which in our case will be $|c_\nu|$ and $\arg\{c_\nu\}$. This is achieved by ``aligning" the elliptical structure of the deterministic autoregression and the stochastic noise. The appeal of reducing to five parameters is that the process will then have the same number of free parameters as an improper order one linear model~\eqref{eq:doublywhite}, but will produce elliptical oscillations using widely linear forcing ({\em cf.} Fig.~\ref{FigNew1}). The order two process for elliptical oscillations of~\cite{rubin2008kinematics} in general requires seven real-valued parameters.

To reduce our model to five free parameters, we equate the process to a bivariate process with elliptical covariance structure. We start from the complex-proper autoregressive process of order one~\eqref{eq:CAR}. We rewrite this process as a bivariate process in terms of $(X'_t \  \, Y'_t)^T$, where $Z'_t =X'_t+\mathrm{i}Y'_t$,
\begin{equation}
\begin{pmatrix}X'_t\\Y'_t \end{pmatrix} = a R \begin{pmatrix}X'_{t-1}\\Y'_{t-1} \end{pmatrix}+\sigma_\epsilon\begin{pmatrix}\epsilon_{1,t}\\\epsilon_{2,t}\end{pmatrix}, \label{eq:bivCAR}
\end{equation}
where,
\[ R=\begin{pmatrix}\cos\theta & \matminus\sin\theta\\\sin\theta & \cos\theta\end{pmatrix}. \]
It follows that \eqref{eq:bivCAR} is an isotropic real-valued bivariate vector process.
Here $\epsilon_{1,t}$ and $\epsilon_{2,t}$ are i.i.d. Gaussian random variables with mean 0 and variance 1. The matrix $R$ accomplishes a rotation by the angle $-\pi<\theta\le\pi$ in the Cartesian plane. For identifiability we set $a\ge 0$ and for stationarity we require $a<1$, as we show in Appendix~\ref{S:Append1}. We then construct a new process $(X_t \ Y_t)^T$, which we call the {\em elliptical bivariate autoregressive process (of order one)}, using an elliptical transformation
\begin{equation}
\begin{pmatrix}X_t\\Y_t \end{pmatrix} =  QP\begin{pmatrix}X'_t\\Y'_t \end{pmatrix}, 
\label{eq:bivICAR}
\end{equation}
where
\[Q=\begin{pmatrix}\cos\psi & \matminus\sin\psi\\\sin\psi & \cos\psi\end{pmatrix}, \; P=\begin{pmatrix}\frac{1}{\rho}  & 0\\0 & \rho\end{pmatrix},\]
in which $0\le\psi<\pi$ defines the orientation of an ellipse, and $0<\rho\le 1$ defines the eccentricity, $0\le \varepsilon < 1$, where $\varepsilon= \sqrt{1-\rho^4}$. We call this transformation ``elliptical" as it first stretches the $X$-axis by a factor of $1/\rho$, and compresses the $Y$-axis by a factor of $\rho$, before then rotating these axes through angle $\psi$. This transformation leads to elliptical statistical properties, and the process remains stationary for $a<1$. This can be seen by examining the $2\times2$ covariance matrices associated with~\eqref{eq:bivCAR} and \eqref{eq:bivICAR} which are respectively given by
\[
\E\left\{\begin{matrix}{X'_t}^2 & X'_tY'_t\\X'_tY'_t & {Y'_t}^2 \end{matrix}\right\}=\left(\frac{\sigma^2_\epsilon}{1-a^2}\right)I,\]
and \[
\E\left\{\begin{matrix}X_t^2 & X_tY_t\\X_tY_t & Y_t^2 \end{matrix}\right\}=\left(\frac{\sigma^2_\epsilon}{1-a^2}\right)QP^2Q^T,
\]
where $I$ is the identity matrix. The covariance matrix of $(X'_t \ \, Y'_t)^T$ is isotropic or circular, whereas in general the covariance matrix of $(X_t \ Y_t)^T$ is elliptical with orientation $\psi$ and ratio of semi-minor to semi-major axis $\rho^2$. 

The elliptical bivariate autoregressive process of order one is defined by five parameters, namely $\{a,\theta,\rho,\psi,\sigma^2_\epsilon\}$. We now express the widely linear complex autoregressive process of order one as $Z_t=X_t+\mathrm{i}Y_t$, and in Proposition~\ref{Prop1} we relate the parameters of this process to those of the elliptical bivariate autoregressive process to form a five-parameter process.

\begin{theorem}\label{Prop1}
Suppose the process $(X_t \ Y_t)^T$ is an elliptical bivariate autoregressive process of order one, as defined in~\eqref{eq:bivCAR} and \eqref{eq:bivICAR} by the parameters $\{a,\theta,\rho,\psi,\sigma^2_\epsilon\}$. This process is equivalent to a widely linear complex autoregressive process of order one~\eqref{eq:ICAR}, specified by $Z_t=X_t+iY_t$ where
\begin{multline}
Z_t = a\left\{\cos\theta+\frac{\mathrm{i}\sin\theta}{2}\left(\frac{1}{\rho^2}+\rho^2\right)\right\}Z_{t-1}\\+\frac{a\sin\theta}{2}\left(\frac{1}{\rho^2}-\rho^2\right)e^{\mathrm{i}\left(2\psi-\frac{\pi}{2}\right)}Z^*_{t-1}\nonumber\\
+\sigma_\epsilon\left\{\frac{e^{\mathrm{i}\psi}}{\rho}\epsilon_{1,t}+\rho e^{\mathrm{i}\left(\psi+\frac{\pi}{2}\right)}\epsilon_{2,t}\right\}.
\end{multline}
The relationship with the parameters $\{\lambda,\alpha,\gamma,\phi,\sigma^2_\nu\}$ in the specification of \eqref{eq:ICAR} is given in Table~\ref{Tab1}, with the final redundant parameter, the relation at lag zero $c_\nu$, specified by
\begin{equation}
c_\nu=\sigma^2_\epsilon\left(\frac{1}{\rho^2}-\rho^2\right)e^{\mathrm{i}2\psi}=\sigma^2_\nu\left(\frac{\gamma}{\lambda\sin\alpha}\right)e^{\mathrm{i}\left(\phi+\frac{\pi}{2}\right)}.
\label{eq:setrel}
\end{equation}
\end{theorem}

\begin{table*}
\setlength{\extrarowheight}{1.2em}
\begin{center}
\caption{\label{Tab1}This table provides a mapping between the parameters of the elliptical bivariate process of order one~\eqref{eq:bivICAR}, and the widely linear complex autoregressive processes of order one~\eqref{eq:ICAR}. We require $\lambda\ge0$, $-\pi<\alpha\le\pi$, $\gamma\ge0$ and $\sigma^2_\nu>0$ for the widely linear complex process, and $a\ge0$, $-\pi<\theta\le\pi$, $0<\rho\le1$ and $\sigma^2_\epsilon>0$ for the elliptical bivariate process. The parameters $\phi$ and $\psi$ are unrestricted in this mapping. The function ${\tt atan2}$ is the four quadrant inverse tangent, ${\tt acos}$ is the inverse cosine function and $\sign$ is the signum function. These functions are chosen to ensure that $\alpha$ and $\theta$ have a one-to-one mapping in the range $(-\pi,\pi]$.}
\begin{tabular}{|l|l|}\hline
Bivariate elliptical to widely linear complex & Widely linear complex to bivariate elliptical \\ \hline

$\lambda=a\sqrt{\cos^2\theta+\frac{\sin^2\theta}{4}\left(\frac{1}{\rho^2}+\rho^2\right)^2}$
&  
$a=\sqrt{\lambda^2-\gamma^2}$ \\

$\alpha = {\tt atan2}\left\{\frac{\sin\theta}{2}\left(\frac{1}{\rho^2}+\rho^2\right),\cos\theta\right\}$ 
& 
$\theta  = \sign(\alpha){\tt acos}\left(\sqrt{\frac{\lambda^2}{\lambda^2-\gamma^2}}\cos\alpha\right)$\\

$\gamma = \frac{a}{2}|\sin\theta|\left(\frac{1}{\rho^2}-\rho^2\right)$
& 
$\rho = \left(\frac{\lambda|\sin\alpha|-\gamma}{\lambda|\sin\alpha|+\gamma}\right)^{1/4}$\\

$\phi = 2\psi-\sign(\theta)\frac{\pi}{2}$ 
& 
$\psi = \frac{\phi}{2}+\sign(\alpha)\frac{\pi}{4}$\\

$\sigma^2_\nu =\sigma^2_\epsilon\left(\frac{1}{\rho^2}+\rho^2\right)$ 
 &
$\sigma^2_\epsilon = \sigma^2_\nu\frac{\sqrt{\lambda^2\sin^2\alpha-\gamma^2}}{2\lambda|\sin\alpha|}$ \\ \hline
\end{tabular}
\end{center}
\end{table*}

The proof of Proposition~\ref{Prop1} is given in Appendix~\ref{S:Append2}.
Therefore in order to represent the elliptical bivariate process~\eqref{eq:bivICAR} in terms of the widely linear complex process~\eqref{eq:ICAR}, we have required to use only five parameters. We can therefore simply fix $c_\nu$ using either $\{a,\theta,\rho,\psi,\sigma^2_\epsilon\}$ or $\{\lambda,\alpha,\gamma,\phi,\sigma^2_\nu\}$, as given in~\eqref{eq:setrel}. This equivalence can be verified using the transformations in Table~\ref{Tab1}. We note that this is how the value of $c_\nu$ was set earlier in Fig~\ref{FigNew1}.

The eccentricity parameter, $\varepsilon$, can also be found in terms of the widely linear complex process parameters using Table~\ref{Tab1}
\begin{equation}
\varepsilon=\sqrt{1-\rho^4}=\sqrt{\frac{2\gamma}{\lambda|\sin\alpha|+\gamma}}.
\label{eq:ecc}
\end{equation}
Therefore for the widely linear process to return valid values for the eccentricity, $0\le\varepsilon<1$, we require
\begin{equation}
\gamma\le\lambda|\sin\alpha|.
\label{eq:ineq}
\end{equation}
Additionally, for the widely linear process to be stationary we require $a=\sqrt{\lambda^2-\gamma^2}<1$. Therefore from~\eqref{eq:ineq} it follows that $a>0$. Then we see that stationarity is guaranteed when $\lambda<1$, and otherwise for stationarity we require that
\begin{equation}
\gamma>\sqrt{\lambda^2-1},\quad\textrm{when }\lambda\ge1.
\label{eq:ineq2}
\end{equation}
These inequalities also ensure that $\theta$ and $\sigma^2_\epsilon$ return valid values when mapping parameters from the complex to bivariate specifications. Increasing $\gamma$ increases the eccentricity, until eventually larger values of $\gamma$ are not valid. This means that there is a range of values for $\gamma$, which depends on both $\lambda$ and $\alpha$, for our five-parameter process to be valid and stationary. Combining the inequalities in~\eqref{eq:ineq} and~\eqref{eq:ineq2} we see that we require $\lambda^2<1/\cos^2\alpha$ for our process to be valid and stationary. When $\alpha=0$, the case of no spin, then $\gamma=0$ from~\eqref{eq:ineq} and we require $\lambda<1$ for stationarity. However as $\alpha$ increases then interestingly, there are parameter values for which $\lambda>1$ and the process can still be stationary, unlike the proper case, although this then requires a non-zero $\gamma$ as can be seen from \eqref{eq:ineq2}.

We gain further insight by analyzing the relationships in Table~\ref{Tab1}. The first three parameters on each side of the table, $\{\lambda,\alpha,\gamma\}$ and $\{a,\theta,\rho\}$, have a direct one-to-one mapping, where $\{\lambda,\alpha\}$ and $\{a,\theta\}$ become identical as $\rho\rightarrow1$ or $\gamma\rightarrow0$. The parameters $a$ and $\lambda$ are monotonic functions of each other, as are $\alpha$ and $\theta$, which have the same sign in the range $(-\pi,\pi]$. The bivariate ellipse orientation $\psi$ and the complex-conjugate spin parameter $\phi$ are directly related, but are adjusted depending on the sign of $\theta$ and $\alpha$. The ratio of the variance parameters, $\sigma^2_\nu$ and $\sigma^2_\epsilon$, is determined by the eccentricity. The effect of each parameter can therefore be closely related row-by-row in Table~\ref{Tab1}. 

By connecting to a bivariate process, we have gained the advantages of both specifications: we benefit from the compactness and applicability of a complex representation, and we benefit from the interpretability and physical understanding gained from a bivariate representation. A particularly useful feature of complex signals is that we can perform hypothesis tests for impropriety \cite{schreier2006generalized}, and we will demonstrate the insight gained from such an analysis in our seismic data example in Section~\ref{S:seismic}.

\section{Covariance and Relation sequence}\label{S:autocov}
In this section we compute the covariance and relation sequences for the widely linear complex autoregressive process of order one. It follows directly from~\eqref{eq:ICAR} that the process is Gaussian, as it is a linear combination of complex-valued Gaussian random variables. Therefore the covariance and relations sequences fully specify the process. These sequences would have complicated expansions if expressed analytically, so instead we find recurrence relationships between the lags. First we find that the variance ($\sigma^2_Z$) and relation at lag zero ($c_Z$) are given by
\begin{align}
\sigma^2_Z&=\E\{Z_t Z_t^\ast\}\nonumber\\&=\E\left\{\left(\lambda e^{\mathrm{i}\alpha}Z_{t-1}+\gamma e^{\mathrm{i}\phi}Z_{t-1}^\ast+\nu_t\right)\right.\times\nonumber\\ & \quad\quad\quad\quad\quad\quad\quad\quad \left.
\left(\lambda e^{-\mathrm{i}\alpha}Z_{t-1}^\ast+\gamma e^{-\mathrm{i}\phi}Z_{t-1}+\nu_t^\ast\right)\right\}\nonumber\\
&= (\lambda ^2+\gamma^2) \sigma^2_Z+\lambda \gamma e^{\mathrm{i}(\alpha-\phi)} c_Z+ \lambda \gamma e^{\mathrm{i}(\phi-\alpha)} c_Z^\ast+\sigma^2_{\nu},\label{eq:var0}
\end{align}
\begin{align}
c_Z&=\E \{Z_t Z_t\}\nonumber\\&=\E\left\{\left(\lambda e^{\mathrm{i}\alpha}Z_{t-1}+\gamma e^{\mathrm{i}\phi}Z_{t-1}^\ast+\nu_t\right)\right.\times\nonumber\\ &  \quad\quad\quad\quad\quad\quad\quad\quad \left.
\left(\lambda e^{\mathrm{i}\alpha}Z_{t-1}+\gamma e^{\mathrm{i}\phi}Z_{t-1}^\ast+\nu_t\right)\right\}\nonumber\\
&=2\lambda \gamma e^{\mathrm{i} (\alpha+\phi)} \sigma^2_Z+\lambda ^2e^{2\mathrm{i}\alpha}c_Z+\gamma^2e^{2\mathrm{i}\phi}c_Z^\ast+c_{\nu}.\label{eq:rel0}
\end{align}
We now combine~\eqref{eq:var0} and \eqref{eq:rel0} to solve for $\sigma^2_Z$ and $c_Z$, yielding
\begin{multline}
\begin{pmatrix}\sigma^2_Z\\c_Z\\c_Z^\ast\end{pmatrix}=
\begin{pmatrix}
\lambda ^2+\gamma^2 & \lambda \gamma e^{\mathrm{i}(\alpha-\phi)}& \lambda \gamma e^{\mathrm{i}(\phi-\alpha)}
\\ 2\lambda \gamma e^{\mathrm{i}(\alpha+\phi)} & \lambda ^2e^{2\mathrm{i}\alpha} & \gamma^2e^{2\mathrm{i}\phi}\\
2\lambda \gamma e^{-\mathrm{i}(\alpha+\phi)} & \gamma^2e^{-2\mathrm{i}\phi} & \lambda ^2e^{-2\mathrm{i}\alpha}\end{pmatrix}\begin{pmatrix}\sigma^2_Z\\c_Z\\c_Z^\ast\end{pmatrix}\\+\begin{pmatrix}\sigma^2_{\nu}\\c_\nu\\c_{\nu}^\ast\end{pmatrix}
\end{multline}
and hence
\begin{equation}
\begin{pmatrix}\sigma^2_Z\\c_Z\\c_Z^\ast\end{pmatrix}=
M^{-1}\begin{pmatrix}\sigma^2_{\nu}\\c_\nu\\c_{\nu}^\ast\end{pmatrix}\label{eq:zerolag},
\end{equation}
where
\[
M=\begin{pmatrix}
1-\lambda ^2-\gamma^2 & -\lambda \gamma e^{\mathrm{i}(\alpha-\phi)}& -\lambda \gamma e^{\mathrm{i}(\phi-\alpha)}
\\ -2\lambda \gamma e^{\mathrm{i}(\alpha+\phi)} & 1-\lambda ^2e^{2\mathrm{i}\alpha}&-\gamma^2e^{2\mathrm{i}\phi}\\
-2\lambda \gamma e^{-\mathrm{i}(\alpha+\phi)} & -\gamma^2e^{-2\mathrm{i}\phi} &1-\lambda ^2e^{-2\mathrm{i}\alpha}\end{pmatrix}.
\]
The analytic form for the inverse matrix $M^{-1}$ in~\eqref{eq:zerolag} is provided as part of the online software available at \texttt{http://} \texttt{ucl.ac.uk/statistics/research/spg/software} and is not included here for space considerations. When simulating signals from the process, in addition to satisfying the inequalities specified in \eqref{eq:ineq}, then for the process to be stationary we require that the first observation is generated from the complex-valued normal distribution with mean zero, variance $\sigma^2_Z$, and relation at lag zero $c_Z$. See Section~\ref{SS:likelihood} for more detail on the complex-valued normal distribution. This is how the signals in Fig.~\ref{FigNew1} have been simulated, and more details on this can be found in the supporting online code.

After computing $\sigma^2_Z$ and $c_Z$ from~\eqref{eq:zerolag}, the covariance sequence, $s_\tau$, and the relation sequence, $r_\tau$, can be found for any $\tau>0$ using the following recurrence relationships
\begin{align}
s_{\tau}&=\E\left\{Z_tZ_{t+\tau}^\ast \right\}\nonumber\\&=\E\left\{Z_t\left(\lambda e^{-\mathrm{i}\phi }Z_{t+\tau-1}^\ast+\gamma e^{-\mathrm{i}\phi}Z_{t+\tau-1}+\nu^*_{t+\tau-1}\right)\right\}\nonumber\\
&=\lambda e^{-\mathrm{i}\alpha}s_{\tau-1}+\gamma e^{-\mathrm{i}\phi}r_{\tau-1},
\label{stau}\\
r_{\tau}&=\E\left\{Z_tZ_{t+\tau} \right\}\nonumber\\&=\E\left\{Z_t\left(\lambda e^{\mathrm{i}\phi }Z_{t+\tau-1}+\gamma e^{\mathrm{i}\phi}Z_{t+\tau-1}^\ast+\nu_{t+\tau-1}\right)\right\}\nonumber\\
&=\lambda e^{\mathrm{i}\alpha}r_{\tau-1}+\gamma e^{\mathrm{i}\phi}s_{\tau-1}.
\label{rtau}
\end{align}
Therefore, after solving for $s_0=\sigma^2_Z$ and $r_0=c_Z$, we iterate to find $s_{\tau}$ and $r_{\tau}$ using~\eqref{stau} and \eqref{rtau}. To find the sequences for negative lags we use the simple relationship
$s_{-\tau}=s_\tau^\ast$ and $r_{-\tau}=r_{\tau}$.

From \eqref{stau} and \eqref{rtau} we see that $\lambda$ and $\alpha$ contribute to the exponential decay of the autocovariance---this is expected as autoregressive processes are short memory. Conversely, $\gamma$ and $\phi$ have a ``flipping" effect on $s_\tau$ and $r_\tau$, where the covariance is dependent on the relation at previous lags and vice-versa. This is a consequence of the widely linear form in~\eqref{eq:ICAR}, where the conjugate action creates an iterative `flip' of the process about the real axis in the complex plane.

\section{Methods for Estimating Parameters}\label{S:inference}
In this section we detail how the parameters of the widely linear complex autoregressive process of order one can be estimated from an observed signal using maximum likelihood. We first resolve the exact form of the likelihood in Section~\ref{SS:likelihood}, and then provide an approximate method in the frequency domain in Section~\ref{SS:whittle}. The latter method has practical advantages in real-world applications, as we shall discuss.
\subsection{Maximum Likelihood}\label{SS:likelihood}

Suppose that $Z_t$ follows a stationary widely linear complex autoregressive process or order one, as specified in~\eqref{eq:ICAR} with parameters satisfying~\eqref{eq:ineq} and \eqref{eq:ineq2}, then the probability distribution of $Z_t$ follows a complex-valued normal distribution. For a general complex-valued normally distributed random variable $z$, we denote its distribution by $\mathcal{N}_\mathbb{C}(\mu,\sigma^2,c)$, with mean $\mu$, variance $\sigma^2$, and relation at lag zero $c$. The probability density function of $z$ is then given by
\begin{multline}
p(z)=\frac{1}{\pi\sqrt{(\sigma^2)^2+|c|^2}}\times\\
\exp\left\{-\frac{1}{2}\begin{pmatrix}z^*-\mu^* & z-\mu\end{pmatrix}\begin{pmatrix} \sigma^2 & c \\ c^* & \sigma^2\end{pmatrix}^{-1}\begin{pmatrix}z-\mu \\ z^*-\mu^*\end{pmatrix}\right\}.\nonumber
\end{multline}
It then follows that the probability distribution of the widely linear complex autoregressive process or order one, denoted $Z_t$, is given by
\[
Z_t \sim \mathcal{N}_\mathbb{C}\left(0,\sigma^2_Z,c_Z\right),
\]
where $\sigma^2_Z$ and $c_Z$ are found using~\eqref{eq:zerolag}. For a given observed signal $z_0,\ldots,z_{N-1}$ from the process $Z_t$, the probability of observing the first value $z_0$ directly follows from the probability distribution of $Z_t$,
\begin{multline}
p(z_0;\bm\theta)=\frac{1}{\pi\sqrt{(\sigma^2_Z)^2+|c_Z|^2}}\times\\
\exp\left\{-\frac{1}{2}\begin{pmatrix}z_0^* & z_0\end{pmatrix}\begin{pmatrix} \sigma^2_Z & c_Z \\ c_Z^* & \sigma^2_Z\end{pmatrix}^{-1}\begin{pmatrix}z_0 \\ z_0^*\end{pmatrix}\right\},\nonumber
\end{multline}
where $\bm\theta=\{\sigma^2_Z,c_Z\}$. Next we make use of the Markovian property of the process to find the conditional distribution of $Z_t$ given $Z_{t-1}=z_{t-1}$, for $1\leq t\leq N-1$
\[
(Z_t|Z_{t-1}=z_{t-1}) \sim \mathcal{N}_\mathbb{C}\left(\lambda e^{\mathrm{i}\alpha}z_{t-1}+\gamma e^{\mathrm{i}\phi}z^*_{t-1},\sigma^2_\nu,c_\nu\right),
\]
such that the condition probability of observing $z_t$ given $z_{t-1}$ is
\begin{multline}
p(z_t|z_{t-1};\bm\theta)=\frac{1}{\pi\sqrt{(\sigma^2_\nu)^2+|c_\nu|^2}}\times\\
\exp\left\{-\frac{1}{2}\begin{pmatrix}z_t^*-\mu_{z_t}^* & z_t-\mu_{z_t}\end{pmatrix}\begin{pmatrix} \sigma^2_\nu & c_\nu \\ c_\nu^* & \sigma^2_\nu\end{pmatrix}^{-1}\begin{pmatrix}z_t-\mu_{z_t} \\ z_t^*-\mu_{z_t}^*\end{pmatrix}\right\},\nonumber
\end{multline}
where $\bm\theta=\{\mu_{z_t},\sigma^2_\nu,c_\nu\}$ and
\begin{equation}
\mu_{z_t}=\lambda e^{\mathrm{i}\alpha}z_{t-1}+\gamma e^{\mathrm{i}\phi}z^*_{t-1}.
\label{eq:mut}
\end{equation}
The likelihood of observing the signal $z_0,\ldots,z_{N-1}$ is found by evaluating
\[
p(z_0,\ldots,z_{N-1};\bm\theta)=p(z_0;\bm\theta)\prod_{t=1}^{N-1}p(z_t|z_{t-1};\bm\theta).
\]
The log-likelihood (denoted $\ell_t(\bm\theta)$) is therefore
\[
\ell_t(\bm\theta) = \log\left(p(z_0;\bm\theta)\right)+\sum_{t=1}^{N-1}\log \left(p(z_t|z_{t-1};\bm\theta)\right),
\]
which for a widely linear complex autoregressive process of order one is found to be
\begin{multline}
\ell_t(\bm\theta) = -N\log\pi - \frac{N-1}{2}\log\left((\sigma^2_\nu)^2+|c_\nu|^2\right)\\-\sum_{t=1}^{N-1}\frac{1}{2}\begin{pmatrix}z_t^*-\mu_{z_t}^* & z_t-\mu_{z_t}\end{pmatrix}\begin{pmatrix} \sigma^2_\nu & c_\nu \\ c_\nu^* & \sigma^2_\nu\end{pmatrix}^{-1}\begin{pmatrix}z_t-\mu_{z_t} \\ z_t^*-\mu_{z_t}^*\end{pmatrix}\\- \frac{1}{2}\log\left((\sigma^2_Z)^2+|c_Z|^2\right)-\frac{1}{2}\begin{pmatrix}z_0^* & z_0\end{pmatrix}\begin{pmatrix} \sigma^2_Z & c_Z \\ c_Z^* & \sigma^2_Z\end{pmatrix}^{-1}\begin{pmatrix}z_0 \\ z_0^*\end{pmatrix},
\label{max_lik}
\end{multline}
with $\mu_{z_t}$ given in~\eqref{eq:mut}. The optimal parameter choice $\bm{\hat\theta}$ is then found by maximizing the log-likelihood~\eqref{max_lik}
\begin{equation}
\bm{\hat\theta}=\arg\max_{\bm{\theta}\in\bm{\Theta}}\ell_t(\bm{\theta}),
\label{max_lik2}
\end{equation}
  where $\bm{\Theta}$ is the permitted parameter range for $\bm{\theta}$, recalling that for the five-parameter process, $c_\nu$ is specified by~\eqref{eq:setrel}, and that the inequalities~\eqref{eq:ineq} and~\eqref{eq:ineq2} should also be satisfied.

\subsection{Frequency Domain ``Whittle" Likelihood}\label{SS:whittle}
The parameters of the widely linear complex autoregressive process of order one can also be computed in the frequency domain using Whittle's approximation to maximum likelihood \cite{whittle1953estimation}, known as the `Whittle likelihood.' This approximation of the time-domain likelihood is in the frequency domain, and relies solely on applying Fourier Transforms which can be computed in $\mathcal{O}(N\log N)$ operations. We use a bias-corrected form of the Whittle likelihood, which was extended to complex-valued signals in \cite{sykulski2013whittle} and is given by
\begin{equation}
\ell(\bm{\theta})=-\sum_{\omega\in \Omega}  \left\{\log \left(\left|f(\omega;\bm{\theta})
\right| \right)+J^H(\omega)f^{-1}(\omega;\bm{\theta})J(\omega)\right\},
\label{whittle_complex}
\end{equation}
where $\Omega$ is the set of Fourier frequencies used in the estimation, $\bm\theta$ is the unknown parameter vector, and $J(\omega)$ and $f(\omega;\bm{\theta})$ are given by
\begin{eqnarray}\label{eq:DFT}
J(\omega)=\frac{1}{\sqrt{N}}\sum_{t=0}^{N-1}\begin{pmatrix}
Z_t \\
Z^*_t
\end{pmatrix}e^{-\mathrm{i}\omega t},\\ f(\omega;\bm{\theta})=\begin{pmatrix}
\bar{S}(\omega;\bm{\theta}) &\bar{R}(\omega;\bm{\theta}) \\
\bar{R}^\ast(\omega;\bm{\theta}) & \bar{S}(-\omega;\bm{\theta}) 
\end{pmatrix}.
\end{eqnarray}
The vector $J(\omega)$ is the Discrete Fourier Transform for a signal $Z_0,\ldots,Z_{N-1}$, evaluated at the Fourier frequencies. The matrix $f(\omega;\bm{\theta})$ contains the expected periodogram, $\bar{S}(\omega;\bm{\theta})=\E\{|J(\omega)|^2\}$, and the expected {\em complementary} periodogram, $\bar{R}(\omega;\bm{\theta})=\E\{J(\omega)J(\omega)\}$, and is dependent on both the signal length $N$, and the parameter vector $\bm{\theta}$. Expected periodograms are used in~\eqref{whittle_complex}, rather than theoretical spectral densities, as this removes the known bias effects from using Discrete Fourier Transforms in~\eqref{whittle_complex} for finite $N$. We then compute $\bar{S}(\omega;\bm{\theta})$ and $\bar{R}(\omega;\bm{\theta})$ directly from the covariance and relation sequences using the relationships
\begin{align}
\bar{S}(\omega;\bm{\theta})=\sum_{\tau=-(N-1)}^{N-1}s_\tau(\bm\theta)\left(1-\frac{|\tau|}{N}\right)e^{-\mathrm{i}\omega\tau},
\label{eq:blurred}\\
\bar{R}(\omega;\bm{\theta})=\sum_{\tau=-(N-1)}^{N-1}r_\tau(\bm\theta)\left(1-\frac{|\tau|}{N}\right)e^{-\mathrm{i}\omega\tau}.
\label{eq:blurred2}
\end{align}
The expected complementary periodogram $\bar{R}(\omega;\bm{\theta})$ will in general be a complex-valued quantity. The usefulness of~\eqref{eq:blurred} and \eqref{eq:blurred2} is that they can be computed in $\mathcal{O}(N\log N)$ time as they are Discrete Fourier Transforms, ensuring the Whittle likelihood remains an $\mathcal{O}(N\log N)$ procedure in this bias corrected version. This method is proven to be statistically consistent in \cite{sykulski2016debiased}, and is shown to significantly remove bias effects for sample sizes as large as 1,000 data points.

For the widely linear complex autoregressive process of order one, $s_\tau$ and $r_\tau$ in ~\eqref{eq:blurred} and \eqref{eq:blurred2} are computed using the recurrence relationships given in Section~\ref{S:autocov}, which is an $\mathcal{O}(N)$ computation. Both $f(\omega;\bm{\theta})$ and $J(\omega)$ are then computed using Fast Fourier Transforms, thus giving $\mathcal{O}(N\log N)$ efficiency overall. The optimal parameter choice $\bm{\hat\theta}$ is then found by maximizing the Whittle likelihood in the same way as~\eqref{max_lik2}.

The advantage of performing maximum likelihood in the frequency domain is that we can restrict the range of frequencies used in the summation in~\eqref{whittle_complex}. This allows the parameter estimation to be performed {\em semi-parametrically}, by ignoring frequencies that are known to be contaminated or not specified well by the model. For example, this was used in~\cite{sykulski2015lagrangian} to remove the effect of eddies when estimating the parameters of a turbulent flow model for the ocean surface. We will also employ such semi-parametric procedures in our data example in Section~\ref{S:seismic}.

We note that the Whittle likelihood can be alternatively used with tapered spectral estimates in~\eqref{eq:DFT}, where the triangle kernel $1-|\tau|/N$ in~\eqref{eq:blurred} and \eqref{eq:blurred2} is then replaced with a modified kernel of smaller width, as documented in~\cite{sykulski2016debiased}. Tapering the likelihood helps reduce mean square error in parameter estimates when the process is long memory or has steep spectral slopes. As the widely linear complex autoregressive process of order one is short memory, then tapering is unnecessary and we use the periodogram approach defined in~\eqref{whittle_complex}--\eqref{eq:blurred2}.

\section{Application to Seismic Data}\label{S:seismic}
In this section we investigate using the widely linear complex autoregressive process of order one as a model for seismic data. We analyze the seismic trace from the Feb 9th 1991 Solomon Islands earthquake, as presented in Fig.~\ref{Fig1} in Section~\ref{S:intro}. The data is sampled every second and is freely available from \texttt{http://ds.iris.edu/wilber3}. All the results in this section (and all figures in this paper) are exactly reproducible with MATLAB code available from \texttt{http://} \texttt{ucl.ac.uk/statistics/research/spg/software}.

The seismic trace consists of three components: a radial, vertical, and transverse signal \cite{walden1990estimating}. We model the radial and vertical components as a complex-valued signal, as they are strongly coupled due to the presence of a Rayleigh wave. We do not analyze the transverse component as the expression of the Rayleigh wave does not exist in the transverse signal~\cite{lilly2011modulated}. The radial, $X_t$, and vertical, $Y_t$, signal are displayed in Figs.~\ref{Fig1}(a) and~\ref{Fig1}(b) respectively. Our analysis will first focus on the segment between the dashed vertical lines in the figure. We combine these signals within this partition to form a single complex-valued signal, $Z_t=X_t+iY_t$, as displayed on the complex plane in Fig.~\ref{Fig1}(c). The signal has evident improper structure, as can be seen from the elliptical paths of the signal in the complex plane.

\begin{figure}[t]
\centering
\includegraphics[width=3.1in]{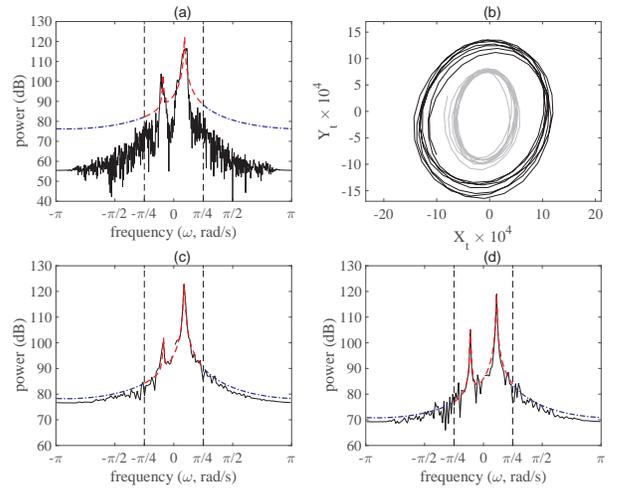}
\caption{\label{Fig2}Panel (a) displays the periodogram ({---}) and model fit ({\color{red}- - -}) of the signal $Z_t$ displayed in Fig.~\ref{Fig1}(c). Panel (b) shows $Z_t$ in the interval (UTC) 16:33:46 to 16:36:26 ({---}) and 16:38:19 to 16:40:59  ({\color{mygray}---}). The periodogram ({---}) and model fit ({\color{red}- - -}) are displayed for $Z_t$ in the intervals (c) 16:33:46 to 16:36:26 and (d) 16:38:19 to 16:40:59. In (a), (c), and (d), parameter estimation is performed for $\omega\in[-\pi/4,\pi/4]$, as indicated by the vertical dashed boundaries, and we extend the fitted lines to all frequencies ({\color{blue}-$\cdot$-}).}
\end{figure}

We first fit the widely linear complex autoregressive process of order one to the entire signal displayed in Fig.~\ref{Fig1}(c) using the Whittle likelihood, as detailed in Section~\ref{S:inference}. The periodogram of the data, and the resulting model fit of the periodogram, are displayed in Fig.~\ref{Fig2}(a). For all parameter estimates in this section, we perform the Whittle likelihood estimation semi-parametrically over a reduced range of frequencies, $\omega\in[-\pi/4,\pi/4]$, as the signal energy is strongly concentrated within this frequency range. For complex-valued signals, the spectrum is defined at both negative and positive frequencies, and will in general be asymmetric. The two peaks of different magnitude on each side of the spectra, at approximately the same frequency, indicate elliptical oscillatory motion. Our fitted process has located the frequency of these peaks, but overall is a poor fit to the periodogram. This is due to the nonstationarity of the signal. Inspecting Fig.~\ref{Fig1}(c) in more detail we can see that the amplitude, eccentricity and orientation of the elliptical oscillations are changing in time. Our model, which is stationary, is not able to capture this variable structure.

To investigate these nonstationary effects we separately analyze two segments of the data, each 161 seconds (or data points) long, spanning the periods 16:33:46 to 16:36:26 and 16:38:19 to 16:40:59 (UTC) respectively. The complex-valued signals corresponding to these time periods are displayed in Fig.~\ref{Fig2}(b). The choice of window length is motivated by the example signal itself, and has been selected such that it is as short as possible (to capture as much time variability as possible), while still being able to robustly estimate all five free parameters. For automated windowing procedures we refer the reader to~\cite{adak1998time}, which is outside the scope of this paper. The periodograms and model fits of each segment are displayed in Figs.~\ref{Fig2}(c) and~\ref{Fig2}(d). The process is now seen to be a good fit to these shorter signals. The optimal parameters are significantly different for each segment. For example, the eccentricity estimate in the first segment is $0.29$ whereas in the second segment it is $0.56$. These differing eccentricity estimates are related to the different ratios of the amplitudes of the two peaks in each respective periodogram.

The appropriateness of our process to modeling shorter segments of this signal suggests that a {\em locally stationary} modeling assumption should be used, see e.g. \cite{fuentes2002spectral}. To investigate this in more detail we perform the model fit to a rolling 161-second window over the entire signal.

\begin{figure}[t]
\centering
\includegraphics[width=3.4in]{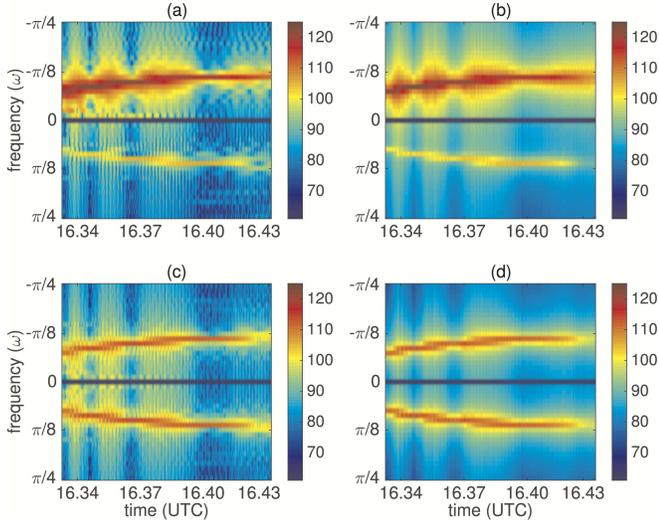}
\caption{\label{Fig3}Spectrograms using a 161-second sliding window of the seismic signal of Fig.~\ref{Fig1}. Panel (a) is the evolving periodogram of $Z_t$, (b) is the evolving model fit to the periodogram using our widely linear process, (c) is the magnitude of the evolving complementary periodogram of $Z_t$ and (d) is the magnitude of the evolving model fit to the complementary periodogram. The color scale is given in decibels.}
\end{figure}

In Fig.~\ref{Fig3}(a) and~\ref{Fig3}(b) we plot the spectrogram of the data---that is, a moving window of the periodogram---together with the spectrogram of the expected periodogram from our model fit. These spectrograms are only shown for the frequencies that have been used in the fit. Note that the zero frequency is not included in the fit as we have removed the sample mean for each segment. From the figure it can be seen that the widely linear complex autoregressive process of order one captures the overall shape of the spectrum at each time slice, as well as its evolution over time. In particular, the process has captured the gradually changing frequency of the oscillations.

We also display, in Fig.~\ref{Fig3}(c) and~\ref{Fig3}(d), the time-frequency plots for the magnitude of the complementary periodogram, and the resulting model fit (respectively). The complementary periodogram forms a Fourier pair with the sample relation sequence, and as a consequence complementary periodograms from observed improper processes are expected to exhibit noticeable structure. This structure has been captured well in the model fit, which is important, as the complementary periodogram is where information regarding impropriety---such as the expected orientation of elliptical motion---is contained. The complementary periodogram is complex-valued, but we have not included plots for its phase here, for space saving considerations.

\begin{figure}[t]
\centering
\includegraphics[width=3.3in]{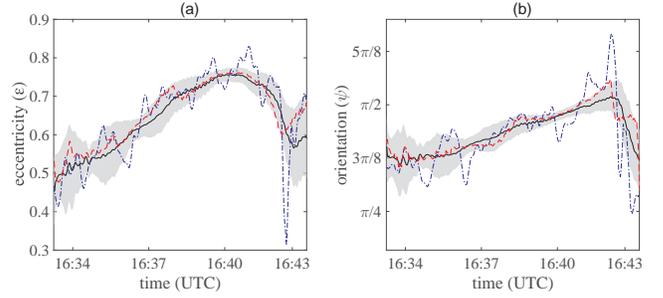}
\caption{\label{Fig4}Estimates of (a) the eccentricity, $\varepsilon$, and (b) the orientation, $\psi$, in radians, of the seismic signal of Fig.~\ref{Fig1}. (----) are the estimates from fitting a widely linear complex autoregressive process of order one across a 161-second sliding window, with the 95$\%$ confidence intervals given in gray. (\textcolor{blue}{-~$\cdot$~-}) are the estimates from the method of~\cite{lilly2010bivariate}, and (\textcolor{red}{- - -}) are the estimates obtained from the Fourier Transform evaluated at the two frequency peaks. All estimates and confidence intervals, including the nonparametric techniques, have been smoothed in time using a moving average window of width 11.}
\end{figure}

In Fig.~\ref{Fig4}(a) and~\ref{Fig4}(b) we display the time-varying eccentricity and orientation parameters from the model fit, calculated using Table~\ref{Tab1} and~\eqref{eq:ecc}. Other informative time-varying summaries can also be found, such as of the noise variance or damping terms. These can be generated as part of the online software. In Fig.~\ref{Fig4} we compare with two alternative methods. First we compare against results obtained from the nonparametric deterministic approach of~\cite{lilly2010bivariate}, which models the signal as a time varying ellipse, thus providing a good comparison to our results despite being a complementary approach. Secondly, we compare against a simple nonparametric approach of comparing the Fourier transform at the positive and negative frequency peaks in the power spectral density, which we denote as $\pm\omega_{\max}$, where it can be readily shown that eccentricity and orientation estimates can be obtained from
\begin{align*}
\hat\varepsilon &= \frac{2\sqrt{|J_Z(\omega_{\max})J_Z(-\omega_{\max})|}}{|J_Z(\omega_{\max})|+|J_Z(-\omega_{\max})|}\\
\hat\psi &= \frac{1}{2}\left[\arg\{J_Z(\omega_{\max})\}+\arg\{J_Z(-\omega_{\max})\}\right],
\end{align*}
with $J_Z(\omega)$ defined as the top row of~\eqref{eq:DFT}. In Fig.~\ref{Fig4}, the estimated eccentricities and orientations broadly agree across the different methods. The values obtained from our parametric model are in general smoother, as the method smooths over information across frequencies when estimating parameters. The usefulness of our stochastic process is that it prescribes a generating mechanism providing physical insight and the ability to replicate signals, which the alternative purely diagnostic metrics do not provide.

Another useful feature of a stochastic modeling approach is that we can calculate confidence intervals for parameter estimates, by numerically computing the Hessian of the Whittle likelihood, as detailed in \cite{sykulski2015lagrangian}. In Fig.~\ref{Fig4} we include the 95\% confidence intervals for our parameter estimates, where care must be taken when assessing significance across time as these are not pointwise simultaneous confidence intervals.

Finally, another advantage of the stochastic modeling approach is that we can perform a parametric hypothesis test for impropriety, to test for statistical significance for when an improper model should be used. This approach is simpler than performing the test of~\cite{schreier2006generalized}, under the assumption that the parametric model is appropriate. We perform the parametric test by also fitting the {\em proper} complex autoregressive order one process~\eqref{eq:CAR} to rolling windows of observations, in exactly the same manner as performed for the widely linear improper process. We then perform a likelihood ratio test, as proposed in \cite{sykulski2013whittle}, to see if there is significant statistical evidence to suggest the null hypothesis of a proper process should be rejected in favor of an improper process. To do this we compute the likelihood ratio statistic, given by $W=2\{\ell(\bm{\theta}_{\rm alt})-\ell(\bm{\theta}_{\rm null})\}$, where ``alt" and ``null" denote the alternative and null models respectively. We compare the likelihood ratio statistic with the 95th percentile of a $\chi^2_2$ distribution. The $\chi^2_2$ distribution is used because there are two additional parameters in the alternate than in the null.

\begin{figure}[t]
\centering
\includegraphics[width=3.3in]{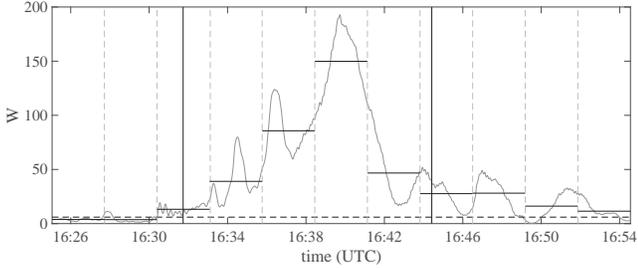}
\caption{\label{Fig5}The likelihood ratio statistic $W$ ({\color{mygray}----}) over time, smoothed with a moving average window of width 11, from fitting the proper~\eqref{eq:CAR} and widely linear improper~\eqref{eq:ICAR} complex autoregressive order one processes to $Z_t$ across a 161-second sliding window. The signal is then divided into 11 non-overlapping windows, as indicated by the vertical gray-dashed lines, where the vertical black solid lines indicate the analysis window of Figs. 4--6. The likelihood ratio statistic for each of these windows is then given by (----). We also display the 95th percentile of a $\chi^2_2$ distribution (- - -), and after applying a False Discover Rate (FDR) procedure to control the rate of false positives, we reject propriety in each segment except the first two.}
\end{figure}

The results of the test are displayed in Fig.~\ref{Fig5}, where we have extended the analysis and computed $W$ over a longer period of time. Similarly to Fig.~\ref{Fig4}, care must be taken here when performing such an analysis over time, as a correction must be made for multiple testing, to control the rate of false positives. As a result, we have divided the analysis into 11 non-overlapping windows, as indicated, and reported the likelihood ratio test statistic within each window. Then to control the rate of false positives, rather than rejecting all segments with $p$-values less than .05 (found using the $\chi^2_2$ distribution), a False Discovery Rate (FDR) procedure is applied using the Benjamini-Hochberg procedure~\cite{benjamini1995controlling}. This procedure ranks the $p$-values in ascending order (denoted $p_1,\ldots,p_{11}$) and finds the largest $j$ such that $p_j\leq.05j/11$, and then rejects all segments corresponding to $p_1,\ldots,p_j$. This procedure formally requires data segments to be independent, and while mild correlations do exist, these can only result in positive correlations between the statistics (as $\chi^2_2$ distributions can only be positively correlated). Therefore we may still employ this procedure, but the rejection rate is conservative.

The FDR analysis, which is included in the online code, rejects all but the first two segments, which have the highest associated $p$-values and are before the arrival of the Rayleigh wave. Propriety is rejected within our main analysis window of Figs. 4--6, and also afterwards where there is still some seismic activity (as can be seen in Fig.~\ref{Fig1}). We can see that the rejection of propriety in favor of our model is most significant at time points where the signal is most eccentric, around 16:40 ({\em cf.} Fig.~\ref{Fig4}(a)), which is intuitive as here a circular/proper model is the least appropriate.

\section{Conclusions}
In this paper we have proposed a widely linear complex autoregressive process of order one. The key novelty of the stochastic process is that impropriety is constructed by relating the process to its conjugate at the previous timestep using a widely linear representation, building on ideas developed in~\cite{navarro2008arma} for higher order ARMA processes. Our approach is in contrast to alternative approaches to modeling improper complex autoregressive processes, where only the noise component is improper. Our stochastic process can generate improper structure in the form of elliptical oscillations, which is not possible using alternative order one processes in the literature. 

We reduced the specification of our process from seven free parameters to five free parameters, by relating the process to a bivariate elliptical process with interchangeable parameters, and ``aligning" the ellipticity of the signal and noise. Reducing to five free parameters has the advantage that parameters of our process are easier to identify and estimate in real-world problems. Furthermore, linking to a bivariate process provided the benefits of using both representations, where we were then able to find conditions for stationarity, and describe the structure of the elliptical oscillations. In general the parameter connections between the representations are non-trivial, but transforming between representations provided useful insight, shedding light on the behavior of the more compact complex widely linear representation.

A promising direction for gaining insight into the full unconstrained seven-parameter widely linear specification, is to relate the process to a bivariate process with two separate elliptical transforms---one each for the autoregressive and noise components in~\eqref{eq:bivCAR}---thus also now being specified by seven parameters. Such an analysis should be performed if the problem is known to have elliptical signal and noise structure that is not aligned.

Another important innovation of this paper is that we have provided time- and frequency-domain techniques to parameter estimation, and then applied them to demonstrate how our widely linear improper process can effectively capture elliptical oscillations in observed seismic traces. An advantage of the complex-valued approach is that frequency-domain understanding then becomes natural, as the asymmetry in the power spectra defines preferred direction of rotation.
The process we propose has the potential to be applied to other improper signals in numerous applications, including for example modeling elliptical eddies as documented in \cite{lilly2006wavelet}, or modeling phase-shifted stochastic cycles in econometric time series~\cite{runstler2004modelling}.

\appendices
\section{Stationarity of the bivariate elliptical autoregressive process}\label{S:Append1}
From~\cite[Ch. 11]{zivot2007modeling} we have that for~\eqref{eq:bivCAR} to be stationary we require that the eigenvalues of the matrix
\[
aR=\begin{pmatrix}
a\cos\theta & \matminus a\sin\theta \\  a\sin\theta & a\cos\theta
\end{pmatrix}
\]
have modulus less than one. The eigenvalues of this matrix are found to be
\[
\lambda_1 = a\cos\theta+\mathrm{i} a\sin\theta,\quad
\lambda_2 = a\cos\theta-\mathrm{i} a\sin\theta.
\]
Therefore as $|\lambda_1|=|\lambda_2|=a$, it follows that requiring both $|\lambda_1|<1$ and $|\lambda_2|<1$ for stationarity is equivalent to requiring that $a<1$ (as we already have that $a\geq 0)$.

\section{Proof of Proposition 1}\label{S:Append2}
Combining~\eqref{eq:bivCAR} and \eqref{eq:bivICAR} we have the relationship
\[
\begin{pmatrix}X_t\\Y_t \end{pmatrix} =QP\left\{ aR \begin{pmatrix}X'_{t-1}\\Y'_{t-1}\end{pmatrix}+ \sigma_\epsilon\begin{pmatrix}\epsilon_{1,t}\\\epsilon_{2,t}\end{pmatrix}\right\},
\]
and then substituting $(X'_{t-1} \ Y'_{t-1})^T$ for $(X_{t-1} \ Y_{t-1})^T$, and using that $Q^{-1}=Q^T$, it follows that
\begin{equation}
\begin{pmatrix}X_t\\Y_t \end{pmatrix} =QP\left\{ aRP^{-1}Q^T \begin{pmatrix}X_{t-1}\\Y_{t-1}\end{pmatrix}+ \sigma_\epsilon\begin{pmatrix}\epsilon_{1,t}\\\epsilon_{2,t}\end{pmatrix}\right\}.
\label{eq:ICARmat}
\end{equation}
To simplify~\eqref{eq:ICARmat} we first define
\[
I=\begin{pmatrix}1 & 0 \\ 0 & 1\end{pmatrix}, \quad J=\begin{pmatrix}0 & \matminus1 \\ 1 & 0\end{pmatrix}, \quad K=\begin{pmatrix}0 & 1\\1 & 0\end{pmatrix}.
\]
It then follows that
\[
PRP^{-1}=\cos\theta I + \frac{\sin\theta}{2}\left(\frac{1}{\rho^2}+\rho^2\right)J-\frac{\sin\theta}{2}\left(\frac{1}{\rho^2}+\rho^2\right)K.
\]
We then use the properties that
\[
QIQ^T=I, \quad QJQ^T=J, \quad QKQ^T=\begin{pmatrix}\matminus\sin2\psi & \cos2\psi\\\cos2\psi & \sin2\psi\end{pmatrix},
\]
which allows for a simple expression for $L=QPRP^{-1}Q^T$ where
\begin{multline*}
L = \cos\theta I + \frac{\sin\theta}{2}\left(\frac{1}{\rho^2}+\rho^2\right)J\\-\frac{\sin\theta}{2}\left(\frac{1}{\rho^2}-\rho^2\right)\begin{pmatrix}\matminus\sin2\psi & \cos2\psi\\\cos2\psi & \sin2\psi\end{pmatrix}.
\end{multline*}
Therefore \eqref{eq:ICARmat} simplifies to
\begin{equation}
\begin{pmatrix}X_t\\Y_t \end{pmatrix} =aL\begin{pmatrix}X_{t-1}\\Y_{t-1}\end{pmatrix}+ QP\sigma_\epsilon\begin{pmatrix}\epsilon_{1,t}\\\epsilon_{2,t}\end{pmatrix}.
\label{eq:ICARmat2}
\end{equation}
To reformulate this in terms of $Z_t=X_t+iY_t$ we use the relationship
\begin{equation}
\begin{pmatrix}X_t\\Y_t \end{pmatrix} =\frac{1}{2}T\begin{pmatrix}Z_t\\Z^*_t \end{pmatrix}, \quad \textrm{where} \quad T = \begin{pmatrix} 1 & 1 \\ \matminus \mathrm{i} & \mathrm{i} \end{pmatrix}.
\label{eq:biv2comp}
\end{equation}
Substituting~\eqref{eq:biv2comp} into~\eqref{eq:ICARmat2} and using that $T^HT=2I$, where the subscript $H$ denotes the Hermitian transpose, after rearranging we see that
\begin{equation}
\begin{pmatrix}Z_t\\Z^*_t \end{pmatrix}=\frac{a}{2}T^HLT\begin{pmatrix}Z_{t-1}\\Z^*_{t-1} \end{pmatrix}+ T^HQP\sigma_\epsilon\begin{pmatrix}\epsilon_{1,t}\\\epsilon_{2,t}\end{pmatrix}.
\label{eq:ICARmat3}
\end{equation}
Expanding~\eqref{eq:ICARmat3} and taking the top row we then get the relationship given in the proposition.

\bibliographystyle{IEEEtran}

%







\end{document}